# Fault Localization Models in Debugging

Safeeullah Soomro[#], Mohammad Riyaz Belgaum[#,1], Zainab Alansari[#,2] and Mahdi H. Miraz[#,3]

[#]*College of Computer Studies, AMA International University Bahrain*
[1]*{s.soomro, bmdriyaz, zeinab, m.miraz}@amaiu.edu.bh*
[1]*Universiti Kuala Lumpur, Malaysia*
[2]*University of Malaya, Malaysia*
[3]*Wrexham Glyndŵr University, UK*

*Abstract*— Debugging is considered as a rigorous but important feature of software engineering process. Since more than a decade, the software engineering research community is exploring different techniques for removal of faults from programs but it is quite difficult to overcome all the faults of software programs. Thus, it is still remains as a real challenge for software debugging and maintenance community. In this paper, we briefly introduced software anomalies and faults classification and then explained different fault localization models using theory of diagnosis. Furthermore, we compared and contrasted between value based and dependencies based models in accordance with different real misbehaviours and presented some insight information for the debugging process. Moreover, we discussed the results of both models and manifested the shortcomings as well as advantages of these models in terms of debugging and maintenance..

## I. INTRODUCTION

Software debugging is considered to be one of the salient phases of software engineering. Since this is a common phenomenon that each human being makes some mistakes at different stages of life, man-made software should not be considered as reliable unless it is thoroughly tested and verified. Testing these anomalies thus is an important part of software System. After development phase of software, testing and debugging should be taken from oracle end. Without these steps, we cannot provide better quality software in today's fast track based world.

Any step of a program performing an intended operation or in an unanticipated way is termed to be a fault. More are the number of faults more will be the development cost of the project. There are different types of faults, some them include System faults, Business Logic faults, Functional faults, Graphical User Interface faults, Behavior faults, Security faults etc. Handling such faults in the software testing plays a very vital role.

Debugging, i.e., identifying as well as correcting errors or faults from software or any sort of computer programmes, involves three (3) different phases:

(1) Fault detection: to identify erratic or unforeseen behaviour
(2) Fault Localization: to identify the origin(s) of the problem.
(3) Repair: to correct the problem by either replacing or modifying the existing code(s) or part(s) of the programme identified during the Fault Localization phase.

All faults may originate during the software development phase and these faults are then corrected through debugging models at designing level and implementation phase. According to oracle's point of view, buggy system produces faulty output with some input/output mixtures of circuits in the sense of software languages. From a software engineering viewpoint, software faults can influence overall project cost, which increases cost after detecting bugs or faults of software development phase for software engineers. So primary goal of every software engineer is to detect, locate and localize faults from software system in the early stages of development phase. This effort reduces cost of maintenance and development of software system. This is the exact point where software debugging is involved.

The paper is organized as follows: we discuss software anomalies and faults classifications in Section 2. The Model based diagnosis technique is presented in Section 3. In Section 4, we present both models and discussions. In Section 5, we present related research.

## II. ANOMALIES OF SOFTWARE

In the IEEE Standard Glossary of Software Engineering [1], the phrase software anomaly describes various issues of software life cycle. The anomalies provide standard method of the classification and appellation. When we discuss about the locating or finding and localizing a bug or an error from a computer program, generally we use both terms an error or a bug.

**Definition 1 (Anomaly).** *According to the IEEE standard classification of software anomalies [1], Anomaly can be interpreted as deviations observed either in documentation or in functionality of any software from previously verified ones or from even reference models.*

The above definition is not bound where the presumption comes from. Below are some words are presenting impressions of anomalies:

**System activity** explains proper way of organized activity for an anomaly was recognized. Some possible listings are coding, analysis, testing and support.

**System life cycle** explains to detect the anomaly in phase. Frame work exists that allow to predict the effect of an anomaly with respect to system life cycle phase it was detested in. Some possible categories are implementation, design, testing, maintenance and accuracy but some other sub-categories are involved.

**Suspected problem** provides unmannered ranking of analogies into the main components of a System. This includes final product itself and final outcome, the analysis and test system, and second partial party products.

**System** describes that how respond the anomaly. Operating system crash, program crash, output problem and input problem are the some specified categories with other sub-categories.

**Repeatability** defines how frequently an anomaly was under observation and if anomaly is reprocreatable.

**Project accuracy** defined that accuracy of system within anomaly is exists. A qualitative assessment of correctness, or freedom from error. A quantitative measure of the magnitude of error from whole system. Project must be accurate with testing and verification after implementation of project.

**Product status** is used to classify how anomaly affect the final product and is dignified between affected, unaffected, usable and drag-gable.

The categories of software anomalies are not limited to software and hardware faults. According to the IEEE standard classification they cover all types of anomalies. In our thesis we are only interested in software anomalies. Which are intentionally connected to the faults of the software system. Here we present some precise definitions according to the IEEE standard Glossary of Software Engineering [1], these are commonly used for the software debugging.

**Mistake** A human action that produces as incorrect result. Like an erroneous action of the software engineer or user. Assume that structure has two data fields, one is last name and other is first name.

Like one wants to access the last name but it access first name so it is a mistake. Unluckily, the term mistake is not used in software debugging field.

**Fault** An incorrect step, process or data definition. Like erroneous action in a computer program. From hardware point of view, fault can be described as a malfunction, break or defect in a device or its component.. For instance, a broken component or a short circuit. Usually doing mistake is a fault. In our example fault is mentioned by accessing the first names instead of last name.

**Failure the state of being unable, for an individual component or a system as a whole, to successfully accomplish the functions or meet the required performance parameters it is designed for..** To simplify, wrong results obtained from any software or programme can be considered as failure. For example after compiling program the variable sum should be equal to 5 but computed once is 20. When fault has been introduced into computer program, a failure is likely to occur. In our example value will be change by the abovementioned fault. Also when we access the first name instead of last name the fault will produce. According to our verification based model computed dependences will change due to this fault with specified ones.

**Error** The deviations of in results i.e. the variation measured in an practically obtained results compared to a hypothetical or abstract ones. For example, if a computed and measured length of distance is 1030 meters while the theoretical distance is 1000 meters, the error will be calculated as the difference between these two values which is 30. Assume that if we have specification (name, lastname) . According to our source code when we access to that the (name− > firstname) so we derived dependences (name, lastname), So here is conflict between specified ones and computed ones, so this an error and defined as dependences error.

As above definition we use IEEE standard definitions [1] for the term mistake, fault, error and failure. In software engineering the term bug has the same meaning as the term fault [2]. Also the above definitions are discussed with examples in [3] for more explanation for the readers. The above classifications and definitions was presented from [1] according to software engineering prospectus and provided the deep information for the readers of the software engineering research community.

## III. MODEL BASED DIAGNOSIS

Model-Based Diagnosis (MBD) is a well-known artificial Technique (AI) for the localization and malfunctioning parts in (mostly physical) systems. In this chapter we briefly describe an overview of Model-Based Diagnosis (MBD) with its fundamental concepts. The definitions of MBD as given in [4, 5] and show how this approach can be used to locate faulty components in a given system depicted in Figure [1].

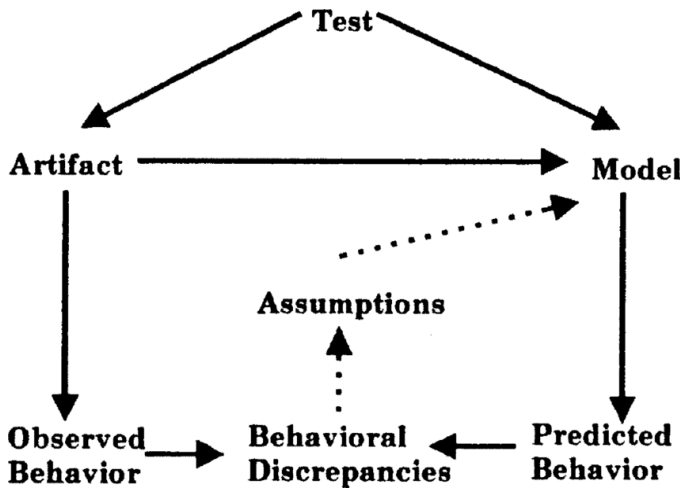

Fig. 1 Model Based Diagnosis [6]

We have used MBD technique is for the software faults particularly value and dependency based models which we are discussing here. Furthermore we are providing comparative analysis to locate and localization of faults of these models. It's a well known technique for the software debugging process where we can work with program specifications and derives computed assumptions to find out the real misbehaviour of the computer programs.

## IV. FAULT LOCALIZATION MODELS AND DISCUSSIONS

In this section we will discuss the value based and dependency based model using some examples to compare the faults according to given technique.

### A. Value Based Model

In this model author [7] introduced faults in the program at value based level. For example if we know the result of the program that out may come 10 from the below code but suddenly output may come different then this fault is very difficult to remove from programs.

Given a = 2 and b = 2;
I.   a = a + 2;
II.  b = b + a
III. c = a + a; // c = a + b;

When we compile the program then output may come 8 which is not right so that MBD technique may help to find out the real misbehaviour of the program according to assertions. So line number III is faulty position of the program. This error will be notified from value based model [7].

### B. Dependency Based Model

This section of the paper manifests abstract dependent fault localization. Aspect System [8] is one of the pioneers to use fault localization technique based on abstract dependencies for fault detection. In fact, the variables used in a programme have inter-relations among themselves. Such relations are more commonly known as abstract dependences. If and only if (*iff*) a new value of "y" may produce a new value of "x", variable "x" is considered to be dependable on variable "y". As for instance, the assignment statement x = y + 1; implies a dependence relation between "x" and "y". Each time the value of "y" is changed, execution of the statement should result in a new value of "x". Further, one more example of such dependencies can be demonstrated by the code segment presented below:

I.   a=b;
II.  b=c;
III. c = a + c;// c = a + b;

In this fragment the program specifications are {(a, b), (a, b), (a,c), (b,c)} whenever introduce fault then computed dependencies are not matched so that Verification Based Model (VBM)[3] may find that fault according to assumptions of all lines and find fault lines. Furthermore In this fragment there are simple assignment statements so we have introduced error in the line number III. The Aspect system is now capable of accepting a program as a input to compute dependencies as well as comparing the measured dependencies with specified dependencies. Thus, if a mismatch is identified comparing the dependencies, the system generates user notifications for bug detection. One of the major limitations of the Aspect system [8] is that it is not able to precisely spot the root-cause(s) of bug thus detected. However, Soomro [3] provided complete information of faulty lines according to model based diagnosis.

According to both models it differs from the finding and localizing faults from programs but both are using Model based Diagnosis technique to overcome faults from the programs in an efficient way.

The author [3] proposed model which may extract computer depencdies from different statements of the program and provide the faulty lines according to dependencies. Whereas the author[7] provide faulty lines of the program according values of the program Both models are providing logical errors of the programs which are very difficult to find in every era of the time. Also very difficult and expensive to maintain softwares now a days and an old days. So these models are playing very important role in accordance to finding real misbehaviours of the programs in an efficient way.

## V. RELATED RESEARCH

The closest match of the work presented in this paper has been identified as the work of Jackson [8]. However, this work focuses on employing abstract dependencies for fault detection rather than localization. Furthermore in [9, 10], for localizing faults, the researchers adopted notion of dependences instead. However, our approach is novel and distinguishes from others: we rather take advantage of using the differences between measured vs. specified dependencies instead of applying variations in the values of the variables detected at any certain line of code. Thus, our approach incorporates the structural properties of program and specification.

Far back in 1999, the authors of [11, 12] developed models for different languages at various abstraction levels in the model-based context. In a broader sense, abstract modeling approaches trade-off details for computational complexity. . In contrast, detailed fault localization capabilities are achieved in the detailed value-level models[,13, 14]. However, the later approach demands considerably large amount of computational power and memory space allocation compared to the earlier one

Another lightweight techniques, commonly known as program slicing, has seen successful application in fault localization [15, 16. The researchers of [12, 16] rather made use of notion of dependences in order to perform fault localization. Thus, the models introduced in [10, 11] suffers from a major shortcoming that they are unable to deal with pre- and post conditions or assertions, in a straightforward way.

In order to handle the faults many researchers have proposed techniques to identify them at an early stage. The authors in [17] have proposed an approach using Genetic Programming to predict the number of faults that may occur in the project and it was validated experimentally using the datasets and has been proved that the proposed model out ruled the existing models.

Various techniques can be adopted during programming itself to prevent the faults like peer review, code analysis being the traditional ones and using the metrics like complexity, fault history. The authors in [18] have conducted a study, to show whether these models can be adopted for vulnerability prediction or some specialized models need to be developed. An empirical study was conducted on the files take from Mozilla Firefox web browser. The results showed that the prediction models existing can be used in place of vulnerability prediction. However, both models required to be improved to reduce the false positives.

TABLE I
COMPARISON OF BOTH MODELS

| S.no | Comparison of Models | | |
|---|---|---|---|
| | Value Based Models | Dependency Based Model | Difference between VBM/DBM |
| 1 | Simple/multiple statements | Simple/Multiple statements | Need program speciation in terms of value and variables |
| 2 | Errors may find from left and right hand side of the code | Only may find fault at the right hand side of the code | VBM works both sides but DBP is limited |
| 3 | Need Specifications in term of values | Need Specification in terms of variable | Both need specifications |
| 4 | Theory of Diagnosis used | Theory of Diagnosis Used | Same Theory used |
| 5 | Faults may find almost from 20K of C/C++ | Faults may find from 20K of Java programs | Different Programming Languages but technique is the same |

In the above Table 1, we have presented both models comparison according to classification of different aspects of finding faults from programs.

## VI. CONCLUSIONS

We have presented here some well known techniques in terms of faults diagnosis. First, we have introduced well known theory and discussed the fault localization models in software debugging area. Our work is solely presented as a simple effort to combine information of verification and values based model in Model based diagnosis which may help readers to understand the importance of theory of diagnosis to locate and localize faults from programs in C/C++ and Java.

Furthermore, we have discussed results and compared both the models which ensure the importance of models in the field of software engineering.

Future work will involve extending the current research towards providing complete information of verification and value based models according to huge experiments which may help the community in terms of software debugging, testing and verification.
.

ACKNOWLEDGMENT

This work is taken from PhD thesis [3]. This work is supported from AMA International University, Kingdom of Bahrain. Authors pay special thanks to Amity International University to publish this work under IEEE proceedings in a Digital Library.

.